\documentclass[a4paper]{jpconf}
\usepackage{graphicx}
\usepackage{amsmath}
\usepackage{times}

\usepackage{citesort}
\def\be{\begin{equation}}
\def\ee{\end{equation}}

\def\etal{\it et al.\,}

\begin{document}
\title{Structure of transition metal clusters: A force-biased 
Monte Carlo approach}
\author{Dil K. Limbu and Parthapratim Biswas}
\address{Department of Physics and Astronomy, The University of 
Southern Mississippi, Hattiesburg, Mississippi 39406, USA}
\ead{dil.limbu@usm.edu}
\ead{partha.biswas@usm.edu (Corresponding author)}

\begin{abstract}
We present a force-biased Monte Carlo (FMC) method 
for structural modeling of transition metal clusters 
of Fe, Ni, and Cu with 5 to 60 atoms. By employing 
the Finnis-Sinclair potential for Fe and the Sutton-Chen 
potential for Ni and Cu, the total energy of the 
clusters is minimized using a method that utilizes 
atomic forces in Monte Carlo simulations.  
The structural configurations of the clusters 
obtained from this biased Monte Carlo approach are 
analyzed and compared with the same from the Cambridge 
Cluster Database (CCD).
The results show that the total-energy of the FMC clusters 
is very close to the corresponding value of the CCD 
clusters as listed in the Cambridge Cluster Database.  A comparison 
of the FMC and CCD clusters is presented by computing 
the pair-correlation function, the bond-angle distribution, 
and the distribution of atomic-coordination numbers in 
the first-coordination shell, which provide information 
about the two-body and three-body correlation 
functions, the local atomic structure, and the bonding 
environment of the atoms in the clusters. 
\end{abstract}

\section{Introduction}
In recent years there has been a rapid development in 
the global optimization techniques, which include state-of-the-art 
evolutionary search strategies to the population-based swarm 
intelligence and differential-evolution approaches~\cite{Engel}. 
In spite of this development, the Monte Carlo (MC) 
methods--based on the Metropolis or related algorithms--continue 
to play a major role in addressing difficult optimization 
problems in science and engineering. Since the calculation of local gradients of 
a potential is computationally more complex than the evaluation of the 
total energy of a system, the MC methods are often preferred 
in many optimization problems where local gradients are 
either not available or computationally too expensive to 
compute.  However, the advantage of the MC methods is often 
outweighed by their slow convergence behavior, which 
require a longer simulation time to produce results with the 
desired accuracy when compared to the Newton-like and
Conjugate-Direction methods.  The main purpose of the 
paper is to explore the usefulness of employing atomic 
forces in Monte Carlo simulations. In particular, we study here 
the application of a simple gradient-based Monte Carlo 
method, originally introduced by Rossky {\etal}\cite{Rossky1978,Allen}, 
to optimize the transition metal clusters of Fe, Ni, and Cu 
and compare the results with the putative global minimum 
of these clusters reported in recent literature~\cite{Elliott2009a,Doye1998}. 

Transition metal clusters (TMCs) have been studied extensively 
from computational and experimental points of 
view~\cite{Elliott2009a,Doye1998,Luo2002,Lopez1996,Parks1995}. 
Theoretical efforts to study TMCs include empirical Monte Carlo and 
molecular-dynamics simulations to semi-empirical tight-binding 
and full {\it ab initio} calculations.  Among numerous 
studies, the structure of Fe clusters obtained from the Finnis-Sinclair 
potential~\cite{Finnis1984,Finnis1986} and that of Ni and Cu clusters bound 
by the Sutton-Chen potential~\cite{Sutton} are of particular 
interest. 
The putative global minima of Ni and Cu clusters of varying sizes 
have been studied systematically by Doye and Wales~\cite{Doye1998} 
using the Sutton-Chen potential. Likewise, Elliott {\etal}\cite{Elliott2009a} 
have recently addressed the computation of the global minima of several 
Fe clusters using the Finnis-Sinclair potential.  In both the cases, 
the authors have employed an improved version of the basin-hopping algorithm of Li and 
Scheraga~\cite{Li1987} using Monte Carlo simulations 
coupled with the Conjugate-Gradient optimization~\cite{Wales1997}.
The total energy and structures of these clusters are 
available from the Cambridge Cluster Database (CCD). 
In the following, we refer these clusters as the CCD clusters and 
use them as a benchmark for the comparison of the total energy 
and structure of Fe, Ni, and Cu clusters obtained from the 
force-biased Monte Carlo (FMC) simulations presented here.

\section{Computational Method}
We began by generating a random configuration consisting 
of N atoms and computing the total energy of the configuration 
using an appropriate interatomic potential. Here, we chose to employ 
the Finnis-Sinclair potential~\cite{Finnis1984} for Fe 
clusters and the Sutton-Chen potential~\cite{Sutton} 
for Ni and Cu clusters. The initial configuration 
was equilibrated at temperature $T$=3000 K for $10^5$ 
Monte Carlo steps (MCS). Thereafter, the temperature 
was decreased by a factor of 0.99 and the system was 
equilibrated for 10$^5$ MCS at each temperature 
until the final temperature of the system was reduced 
to 1 K. The total-energy relaxation was achieved in two steps: 
a) first, we computed the total force on each atom 
in the initial state $n$ and displaced a randomly 
selected atom at site $i$ from an initial state $n$ 
to a proposed state $m$ by~\cite{Rossky1978,Pangali}, 
\be 
{\bf \Delta r_i^{mn}} = \alpha \, {\bf \delta r_i^{mn}} + \beta \, A\, {\bf f_i^n}. 
\label{MC}
\ee
Here, $\beta=\frac{1}{k_B\,T}$ and, $\alpha$ and $A$ are parameters 
that determine the length of the random displacement 
(${\alpha\,\bf \delta r_i^{mn}}$) and the contribution from the 
potential gradient (-${\bf f_i^n}$) at site $i$ in generating 
a proposed configuration $m$, respectively.  The 
displacement ${\bf \delta r_i^{mn}}$ is generally, but not 
necessarily, drawn from a Gaussian distribution with a zero mean and a variance $2A$; 
b) second, the proposed configuration obtained in step (a) 
was either accepted or rejected. Following Rossky 
{\etal}\cite{Rossky1978}, and Allen and Tildesley~\cite{Allen}, 
one can show that the proposed move in Eq.\,(\ref{MC}) can 
be accepted with the probability $P_{mn} = min[1,\exp(-\beta\,\Delta E^{mn}_i)]$, where 
\be 
\Delta E^{mn}_i = \delta E^{mn} + \left[\frac{1}{2}({\bf f_i^n+f_i^m}) \cdot {\bf \Delta r_i^{mn}} 
+ \frac{\beta A}{4}\{({\bf \delta f_i^{mn}})^2 + 2{\bf f_i^n \cdot \delta f_i^{mn}}\}\right]
\label{MC1}
\ee 
\[ 
\delta E^{mn} = E^m -E^n,  \: \quad {\bf \delta f_i^{mn} = f_i^m - f_i^n}. 
\] 
In Eq.\,(\ref{MC1}) above, $E^n$ and $E^m$ are the total energy of the system in the initial 
state and the proposed state, respectively. Likewise, ${\bf f_i^n}$ and 
${\bf f_i^m}$ are the total force on an atom at site $i$ before and 
after the displacement, respectively. In this work, we moved one 
atom at a time but it is straightforward to move a group of atoms 
simultaneously by ensuring that the change of total energy, $\Delta E^{mn}$, 
associated with multi-atom moves, is properly evaluated.  To improve 
the acceptance rate, we adjusted the step length by 
restricting $\alpha$ between 0.001 {\AA} at 1 K and 
0.05 {\AA} at 3000 K. The value of $A$ was chosen in such a way 
that $\beta A \approx$ 4--5 $\times 10^{-3}$ and ${\bf \delta r_i^{mn}}$ 
was a random number drawn from a uniform distribution between -1 and +1. 
In this preliminary study, we made no attempts to optimize the values 
of $\alpha$ and $A$, which can be adjusted during simulations to improve 
the efficiency of the method.  

\section{Results and Discussions}
\subsection{Total energy of Fe, Ni, and Cu clusters}
\begin{figure}[ht]
\begin{center}
\includegraphics[width=50mm,height=31mm]{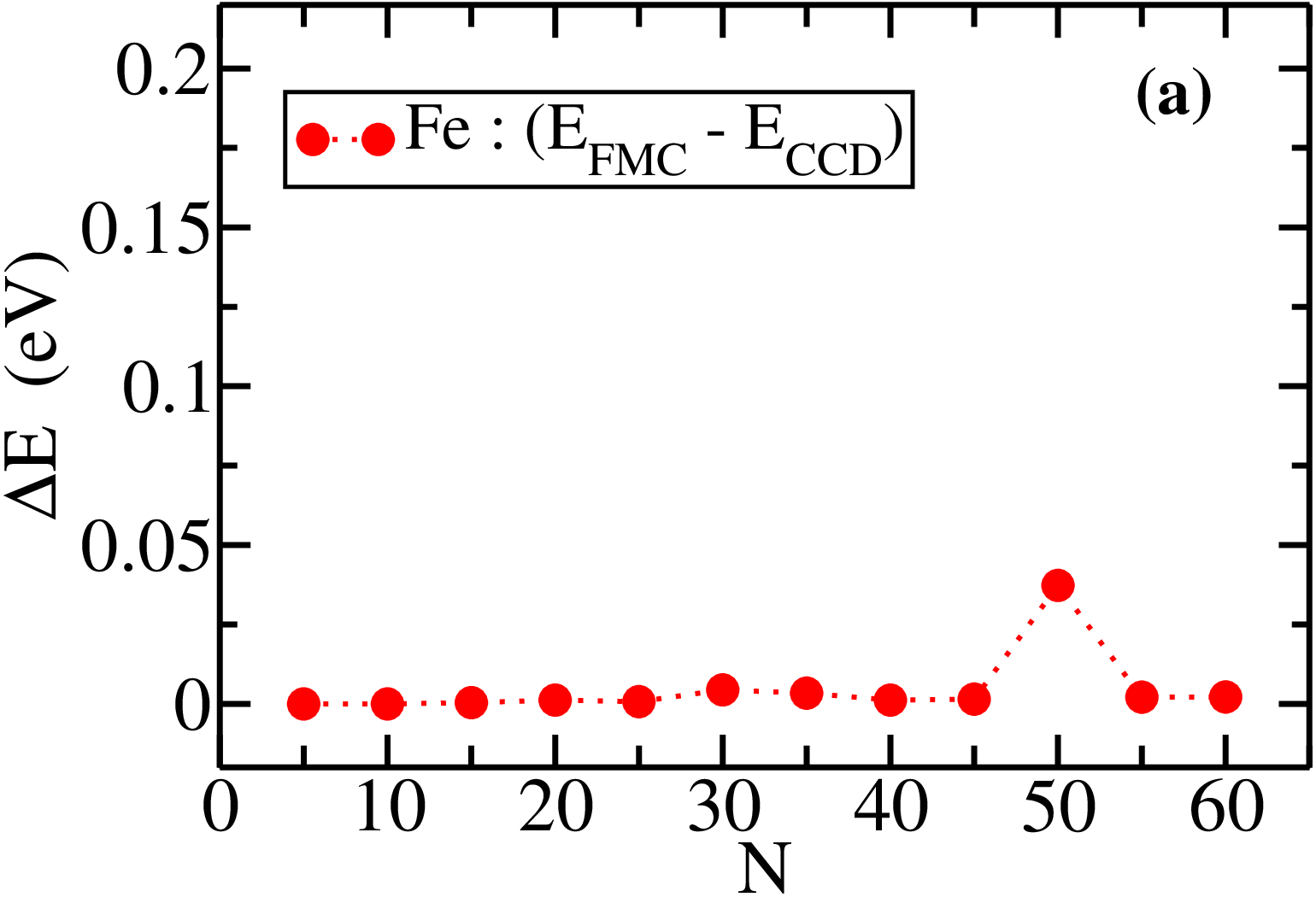}\hspace{2mm}
\includegraphics[width=50mm,height=31mm]{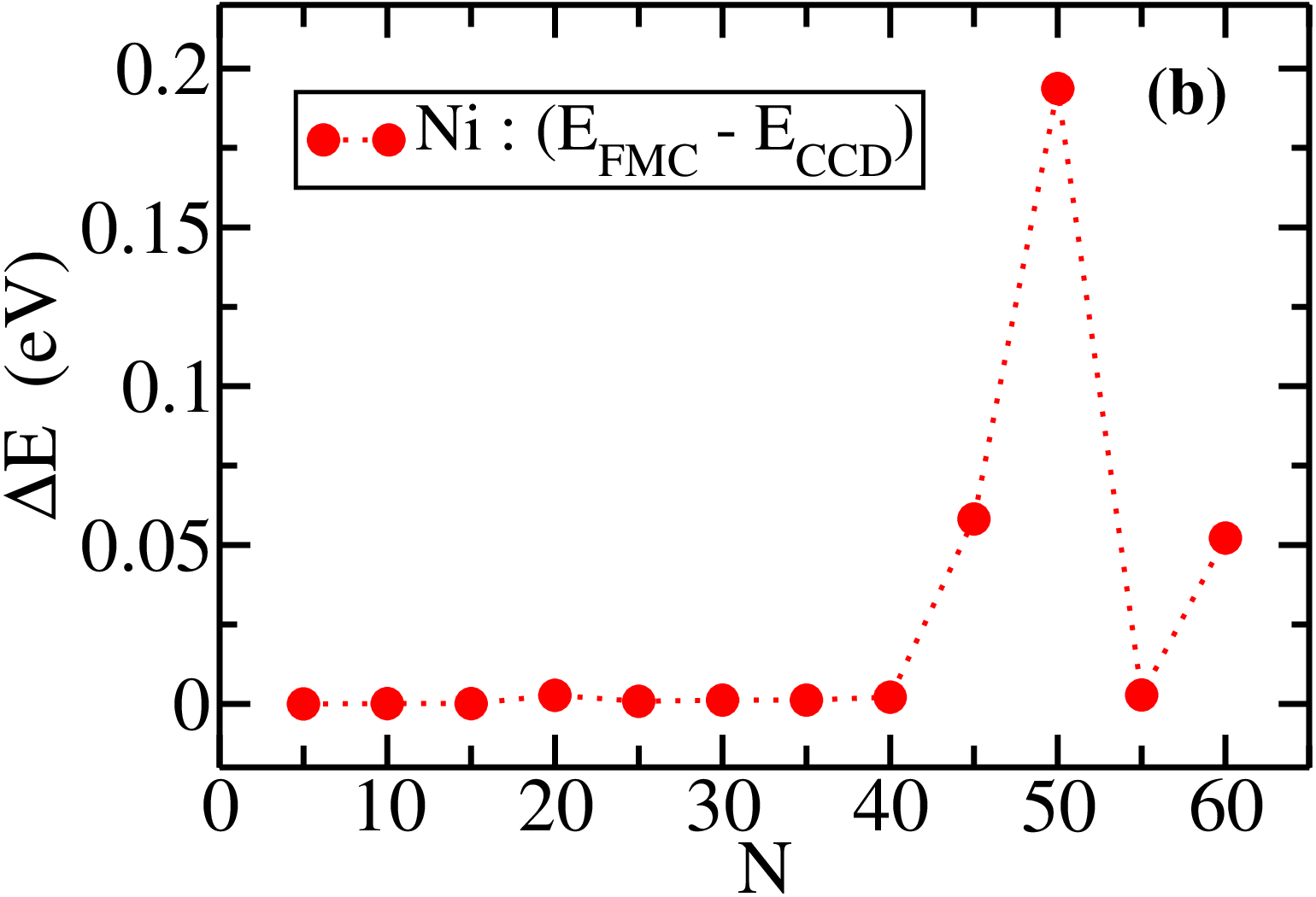}\hspace{2mm}
\includegraphics[width=50mm,height=31mm]{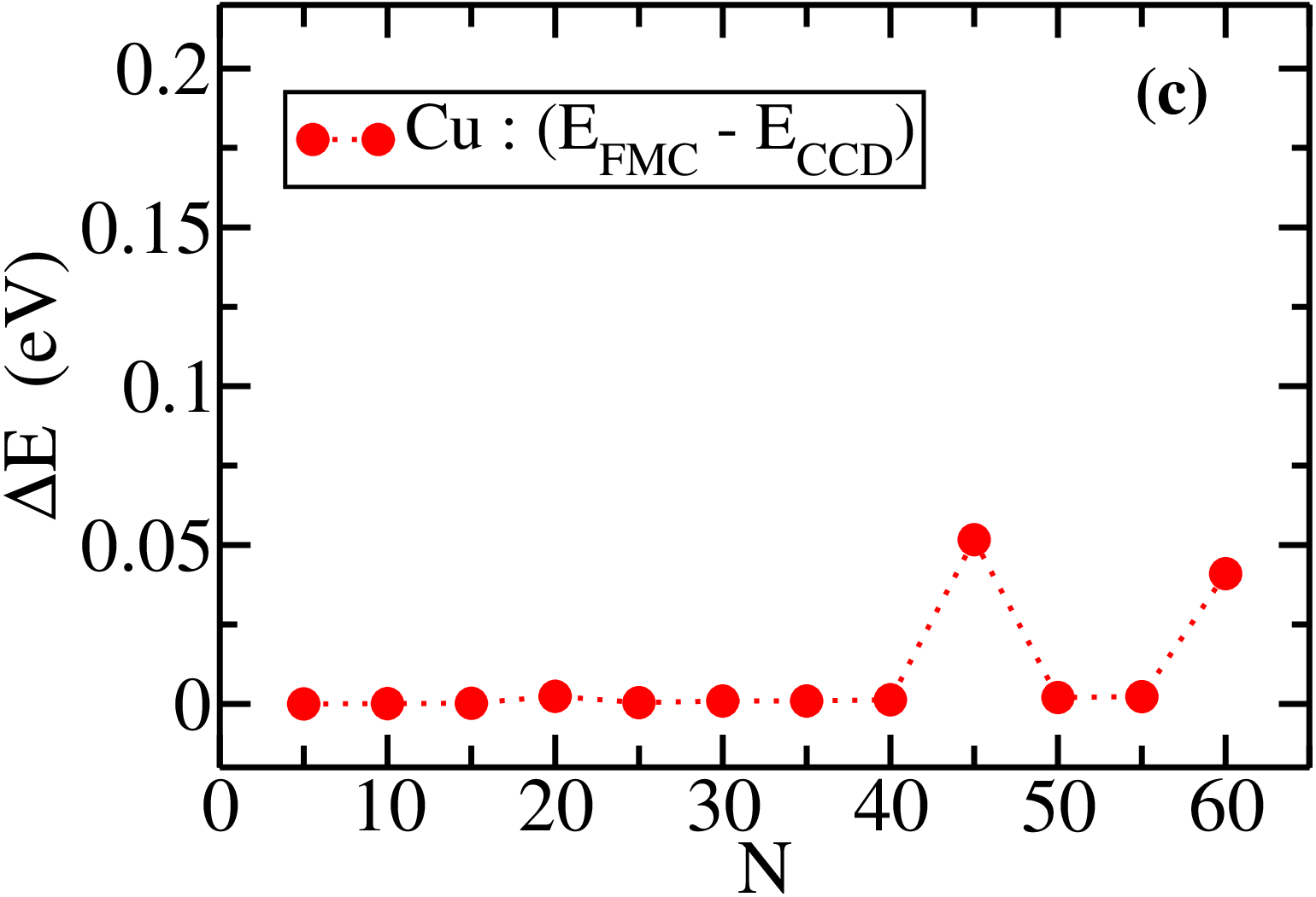}
\caption{\label{ener}
{\small 
Total-energy difference ($\Delta E$) as a function of the 
cluster size ($N$) for a) Fe, b) Ni, and c) Cu clusters. 
}}
\end{center}
\end{figure}

In this section, we present the results from our simulations with 
an emphasis on the structural properties of the clusters. Following 
a comparison of the total-energy values of the FMC clusters with the same 
from the Cambridge Cluster Database~\cite{Elliott2009a,Doye1998}, 
we analyze the two- and three-body correlation functions 
\begin{table}[ht]
\caption{\label{tab_ener} Total energy of Fe, Ni, and Cu clusters from the 
FMC and CCD simulations. 
}
\begin{center}
\begin{tabular}{lccc}
\br
$N$ & Fe (eV) & Ni (eV) & Cu (eV)\\
  &FMC (CCD)&FMC (CCD)&FMC (CCD)\\
 \br
5 & -11.8598 (-11.8598) &-14.6033 (-14.6033) &-11.5120 (-11.5120)\\
10 &-28.5357 (-28.5357) &-32.5487 (-32.5487) &-25.6584 (-25.6585)\\
15 &-46.6371 (-46.6375) &-51.3230 (-51.3231) &-40.4584 (-40.4586)\\
20 &-64.8373 (-64.8385) &-69.9806 (-69.9833) &-55.1663 (-55.1686)\\
25 &-82.9394 (-82.9402) &-89.0701 (-89.0708) &-70.2151 (-70.2155)\\
30 &-101.4469 (-101.4513) &-108.4284 (-108.4296) &-85.4753 (-85.4762)\\
35 &-119.5937 (-119.5971) &-127.9645 (-127.9657) &-100.8758 (-100.8767)\\
40 &-138.5100 (-138.5112) &-147.5979 (-147.6000) &-116.3534 (-116.3547)\\
45 &-156.6982 (-156.6997) &-167.1068 (-167.1649) &-131.7262 (-131.7780)\\
50 &-175.4342 (-175.4720) &-187.0454 (-187.2391) &-147.6007 (-147.6026)\\
55 &-194.6847 (-194.6868) &-207.6107 (-207.6135) &-163.6617 (-163.6640)\\
60 &-214.4256 (-214.4278) &-226.8486 (-226.9008) &-178.8274 (-178.8684)\\
\br
\end{tabular}
\end{center}
\end{table}
and show that the three-dimensional 
distribution of atoms in the FMC and CCD clusters are very close to each 
other as far as these correlation functions are concerned. 
We then address the distribution of the coordination numbers of 
the atoms in order to obtain further information about the atomic arrangement in 
the first-coordination shell of atoms. 
Table \ref{tab_ener} lists the total energy of 
Fe, Ni, and Cu clusters from the FMC simulations. The 
corresponding total energy of the CCD clusters is also indicated 
in Table \ref{tab_ener}. A comparison of the total-energy values suggests that the maximum 
absolute deviation of the total energy (between the FMC and CCD structures) 
is less than 0.2 eV, which translates a percentage deviation 
of 0.1\% (see Ni$_{50}$). Figure \ref{ener} presents the 
total-energy difference, 
$\Delta E=E_{\text{FMC}}$--$E_{\text{CCD}}$, between 
the FMC and CCD clusters.  For Fe clusters of 
up to 45 atoms, the FMC and CCD structures have practically
the same energy, whereas the energy of the remaining few 
clusters with $N>45$ shows a minor deviation of up 
to 0.05 eV from the CCD configuration. A similar observation 
applies for Cu clusters in Fig.\,\ref{ener}(c) and Ni 
clusters in Fig.\,\ref{ener}(b) with the exception 
of Ni$_{50}$, as mentioned earlier. 
It is evident from Table \ref{tab_ener} 
that the optimized FMC clusters are as stable as the 
corresponding CCD clusters as far as the total energy 
of the clusters is concerned.

\subsection{Local structure and atomic environment}
\begin{figure}[ht]
\begin{center}
\includegraphics[width=50mm,height=38mm]{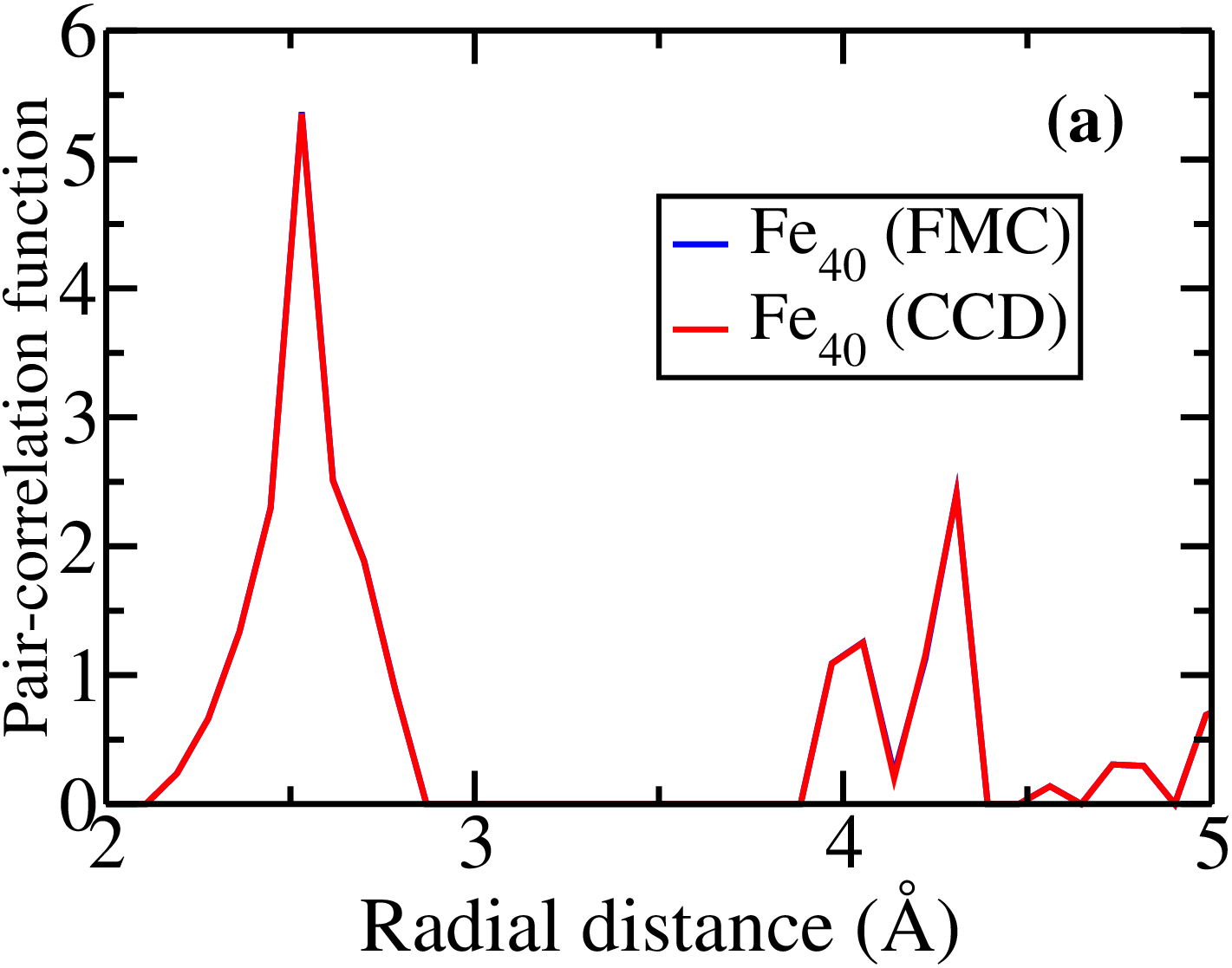}\hspace{2mm}
\includegraphics[width=50mm,height=38mm]{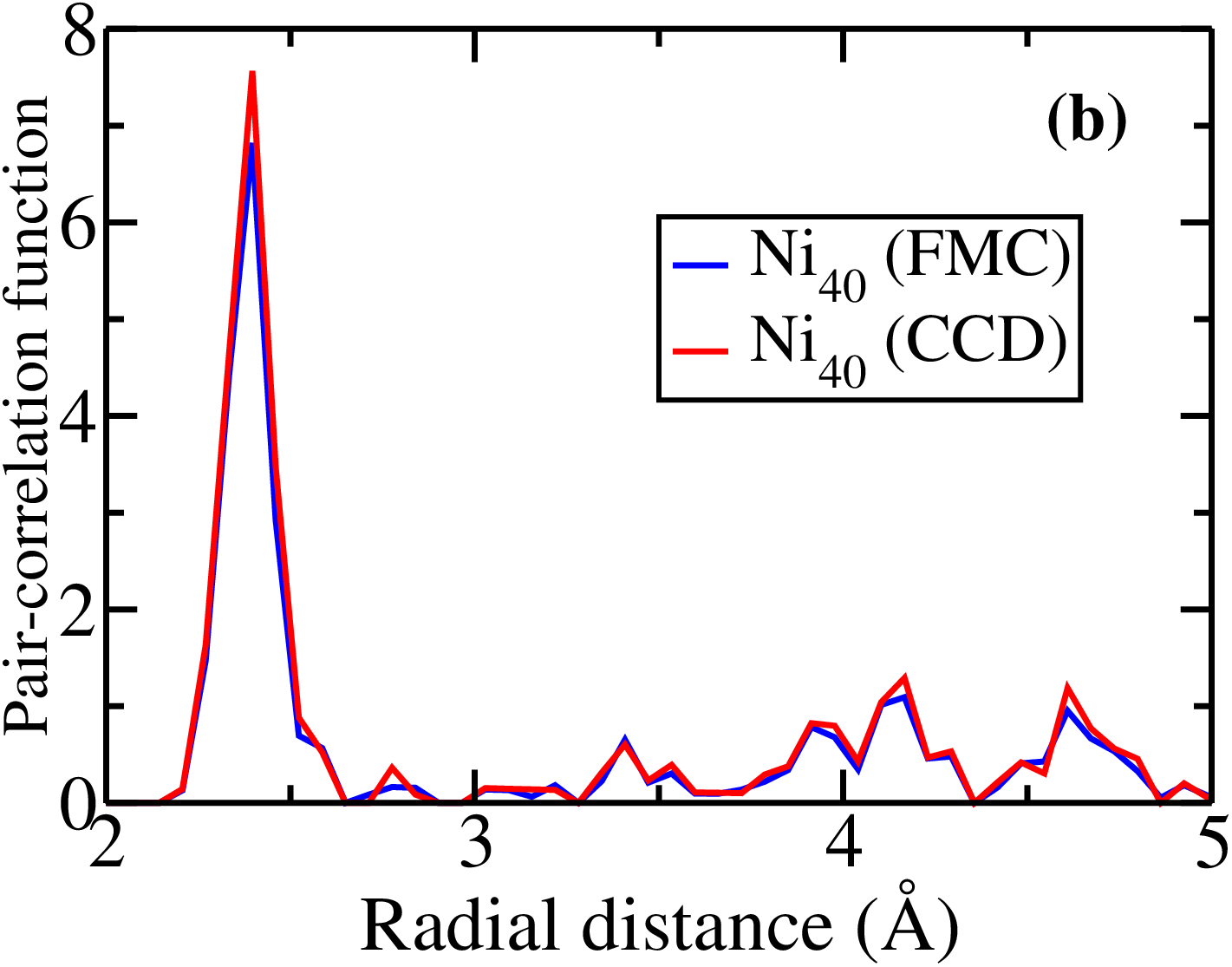}\hspace{2mm}
\includegraphics[width=50mm,height=38mm]{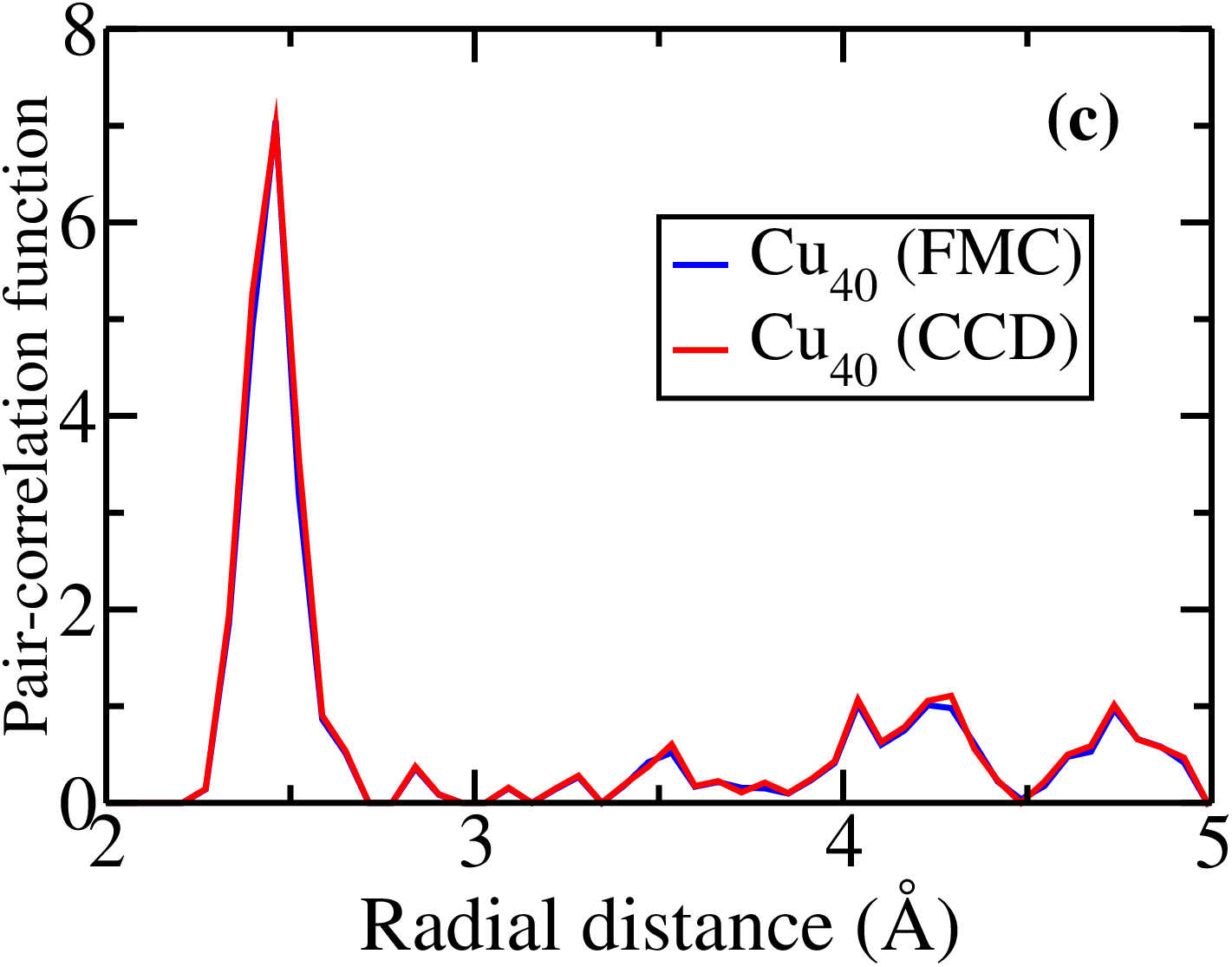}
\caption{\label{rdf} 
{\small 
The pair-correlation functions for a) Fe, b) Ni, and c) Cu clusters 
from the FMC simulations. The corresponding pair-correlation functions 
for the CCD clusters are also shown for comparison. 
}}
\end{center}
\end{figure}

\begin{figure}[ht]
\begin{center}
\includegraphics[width=50mm,height=35mm]{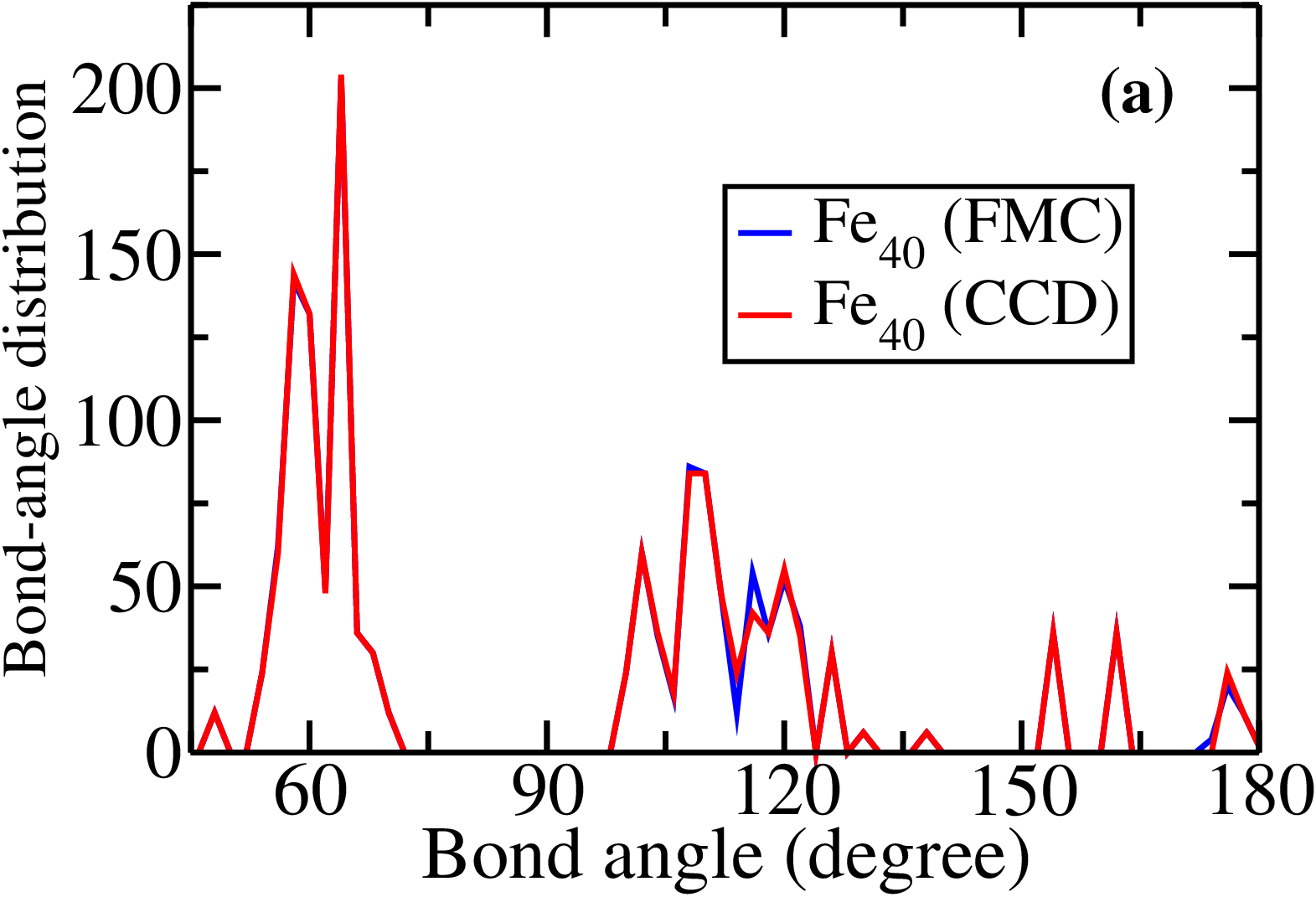}\hspace{2mm}
\includegraphics[width=50mm,height=35mm]{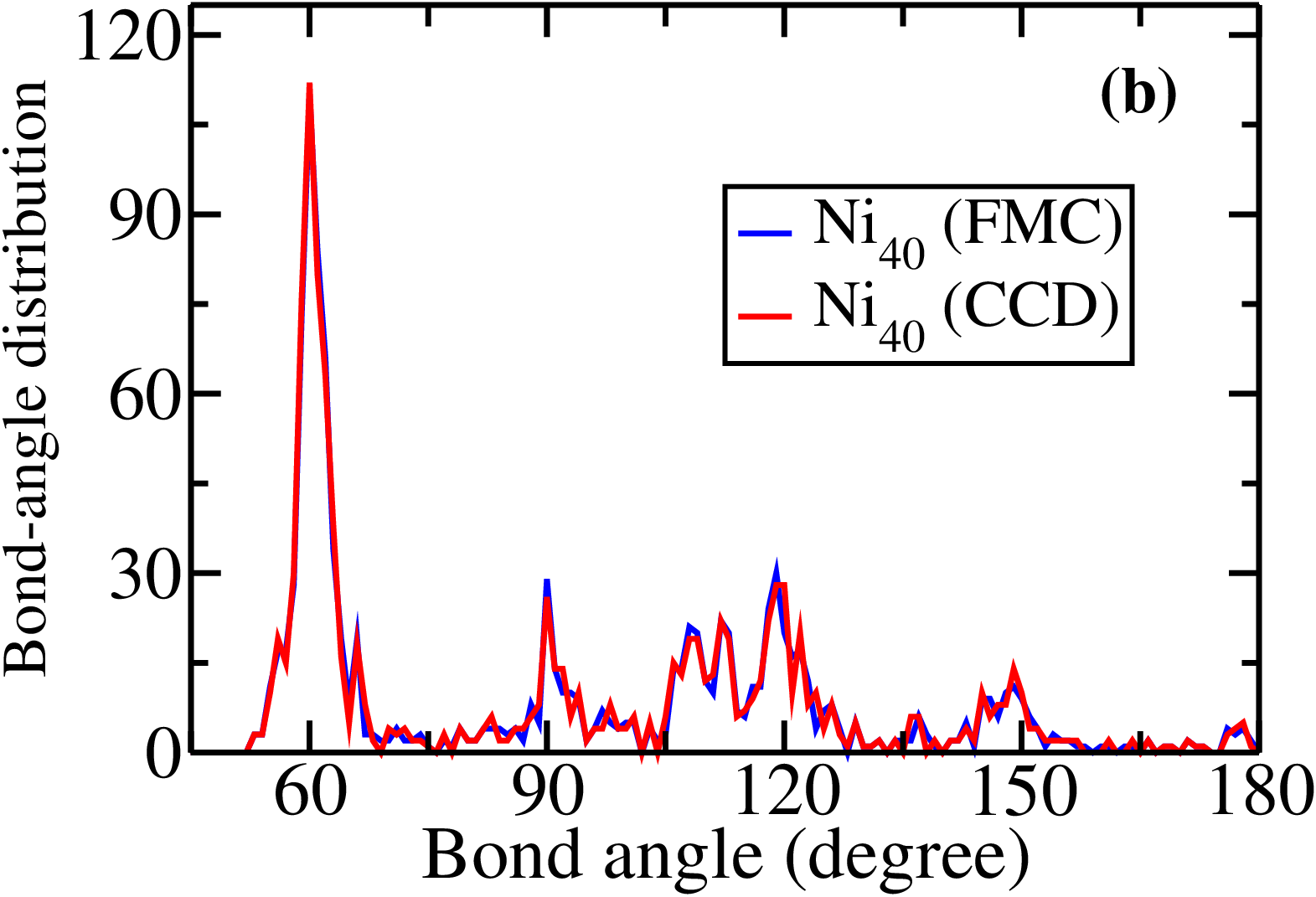}\hspace{2mm}
\includegraphics[width=50mm,height=35mm]{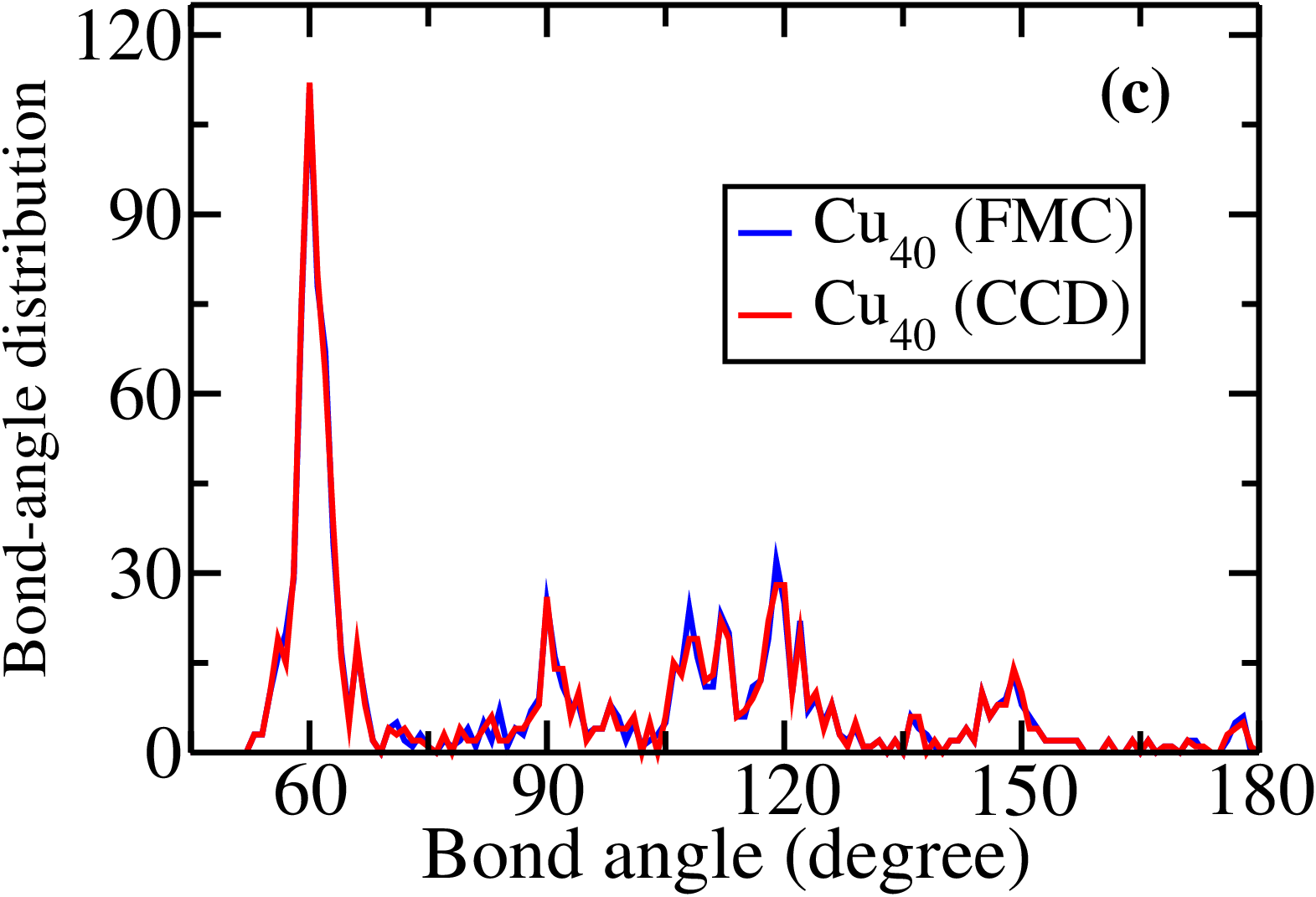}
\caption{\label{bad}
{\small 
Bond-angle distributions for a) Fe, b) Ni, and c) Cu 
clusters from the FMC and CCD structures. 
}}
\end{center}
\end{figure}

We now discuss the two- and three-body correlation functions 
in order to examine the three-dimensional distribution of atoms in the 
clusters. Toward that end, the pair-correlation functions 
for Fe, Ni, and Cu clusters are plotted in Fig.\,\ref{rdf} 
along with the corresponding pair-correlation functions 
for the CCD clusters.  The bond-angle distributions for 
these clusters are also plotted in Fig.\,\ref{bad} for comparison. The results 
indicate that the two-body and three-body correlation 
functions from the FMC structures match very closely with 
their CCD counterparts reflecting structural similarities 
from the point of view of radial and bond-angle distribution 
functions. 

\begin{figure}[ht]
\begin{center}
\includegraphics[width=50mm,height=35mm]{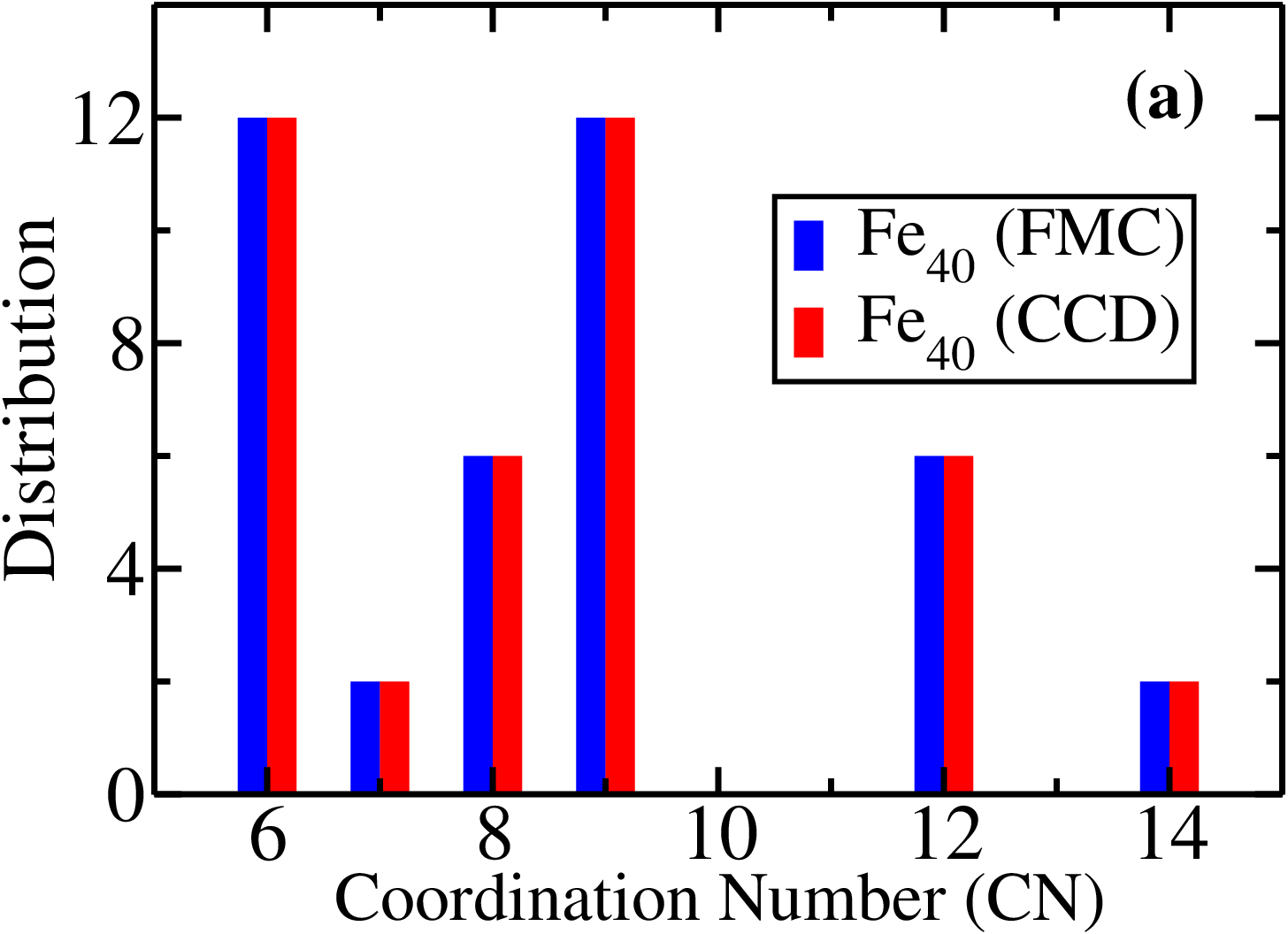}\hspace{2mm}
\includegraphics[width=50mm,height=35mm]{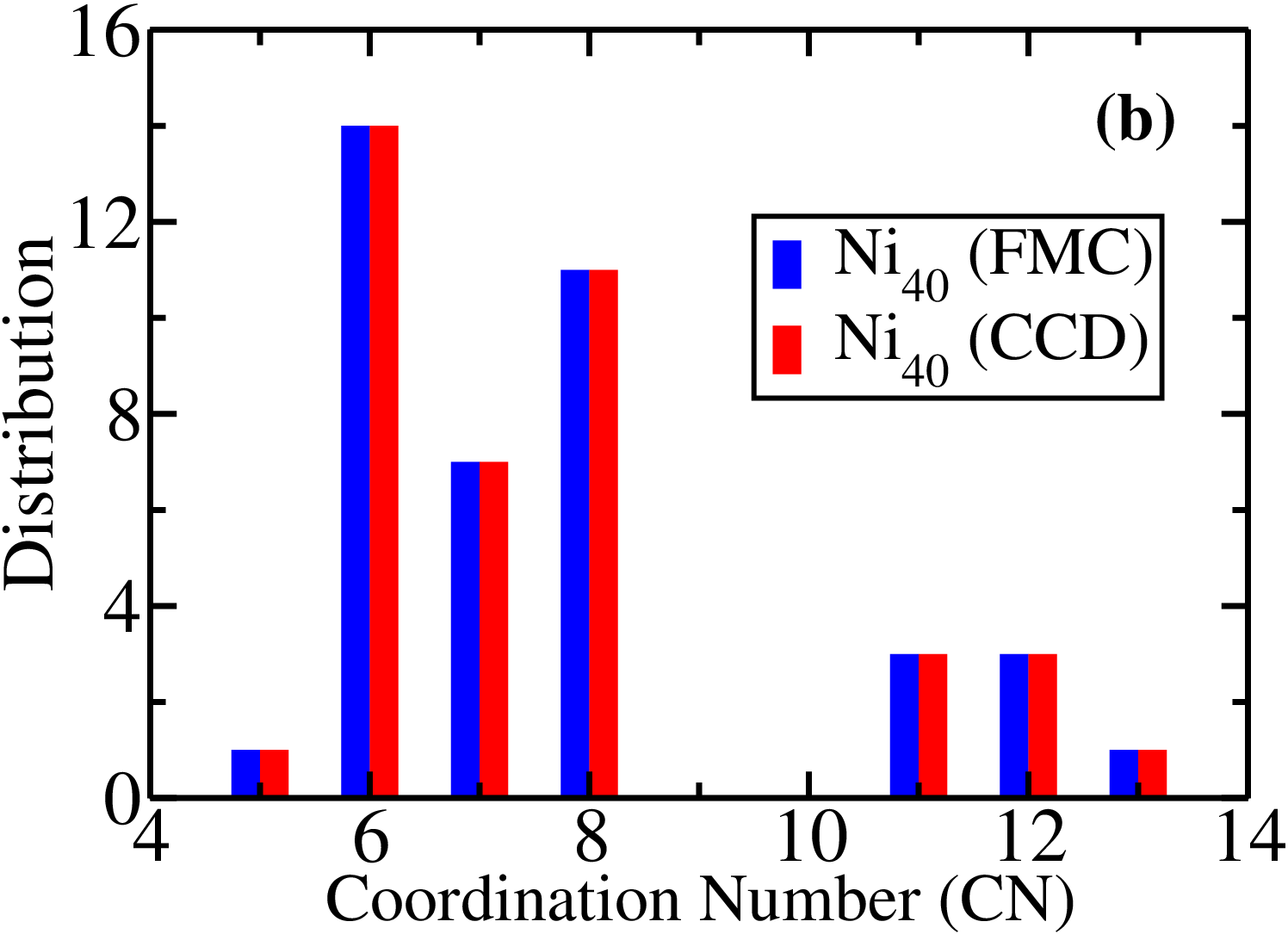}\hspace{2mm}
\includegraphics[width=50mm,height=35mm]{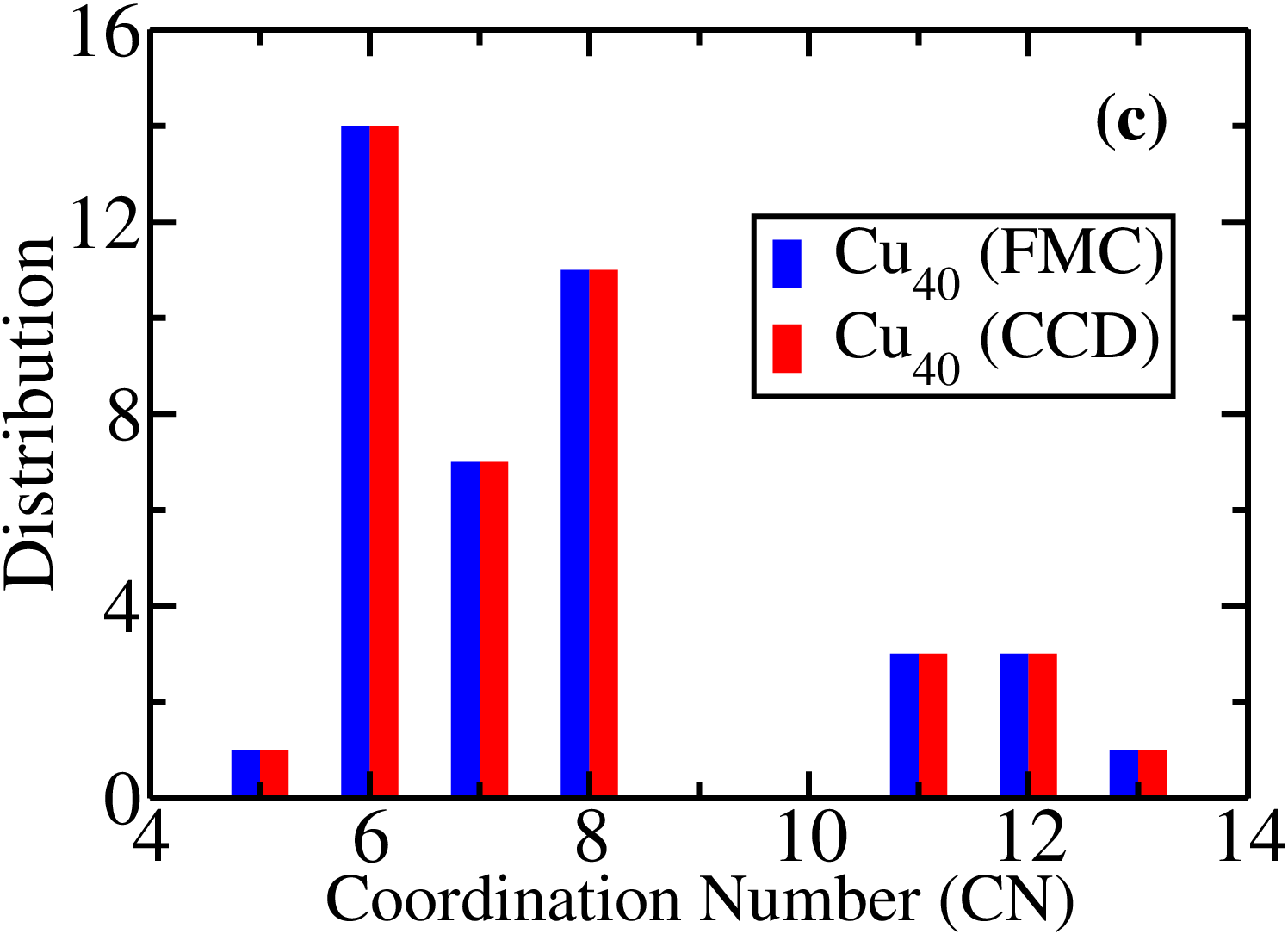}
\caption{\label{CN}
{\small The atomic-coordination numbers in a representative 40-atom a) Fe, b) Ni, and c) Cu cluster. 
}}
\end{center}
\end{figure}
While the pair-correlation function and the bond-angle distribution 
of a cluster provide considerable information about the local atomic 
structure and the bonding geometry of the atoms, these distributions cannot 
uniquely characterize a three-dimensional distribution of atoms in real space. 
Reverse Monte Carlo simulations of amorphous 
materials~\cite{Biswas2004,RMC2001,Biswas2001} -- and the recent work on amorphous 
silicon and amorphous silica~\cite{Biswas2005,Pandey2015,Pandey2016a,Pandey2016b}, using 
hybrid simulation strategies, which employed both the total energy and 
the atomic forces in structural determination of 
complex non-crystalline materials -- indicate that 
one needs to include structural information beyond three-body atomic 
correlations in order to generate a unique atomic arrangement from 
a given set of structural data from experiments. For this purpose, we have computed 
the distribution of the atomic-coordination numbers in 
the first-coordination shell of the atoms and their bonding environment 
in order to obtain additional information on the atomic arrangement of the 
clusters beyond the two-body and the reduced three-body 
correlation functions.  
In Fig.\,\ref{CN}, we have shown the distributions of 
the atomic-coordination numbers for Fe, Ni, and 
Cu clusters.  The results indicate that not only 
the average-coordination number of an FMC cluster and a CCD 
cluster matches with each other, which is evident from the area 
under the first peak of the pair-correlation function 
in Fig.\,\ref{rdf}, but also the full distribution of 
the atomic-coordination numbers as shown in Fig.\,\ref{CN}. 

\section{Conclusion}
In this paper, we have presented a Monte Carlo study of 
total-energy optimization and structural properties of 
transition metal clusters of Fe, Ni, and Cu.  Using a 
force-biased Monte Carlo method, we have optimized the 
total energy of the clusters by employing the 
Finnis-Sinclair potential for Fe and the Sutton-Chen 
potentials for Ni and Cu. 
Unlike conventional Monte Carlo simulations, where the 
acceptance probability of a random displacement of an 
atom is governed by the energy difference between the 
final and initial states only, the present method employs both 
the total-energy difference and atomic forces in determining the 
acceptance probability associated with the Monte Carlo moves. 
The total energy of the clusters obtained from 
the FMC simulations is compared with the data listed in 
the Cambridge Cluster Database (CCD). The results suggest 
that the FMC method can produce structural configurations, which are 
essentially identical to that of the CCD configurations as far 
as the total energy, the pair-correlation function, the 
bond-angle distribution, and the atomic-coordination numbers 
are concerned. The method can be applied to model bulk amorphous solids 
where one is primarily interested in obtaining a set of 
atomic configurations that correspond to a few low-lying 
local minima on the potential energy surface. In a future communication, we shall address 
this problem using an optimized version of the FMC algorithm. 

\ack
The work is partially supported by the National Science 
Foundation (NSF) under Grant No. 1507166. 

\section*{References}
\bibliography{ref9}

\end{document}